\documentclass[twocolumn,showpacs,preprintnumbers,amsmath,amssymb]{revtex4}

\begin{document}
\draft

\title{Friedel Sum Rule for single channel quantum wire}

\author{Swarnali Bandopadhyay and P. Singha Deo}

\address{S. N. Bose National Centre for Basic Sciences, JD Block,
Sector III, Salt Lake City, Kolkata 98, India.}
\date{\today}

\begin{abstract}
Elastic scattering in a quantum wire has several novel features not seen in
$1$D, $2$D or $3$D. In this work we consider a single channel quantum wire
as its application is inevitable in making devices based on quantum
interference effects. We consider a point defect or a single delta function
impurity in such a wire and show how some of these novel features affect
Friedel-sum-rule (FSR) in a way, that is quite unlike in $1$D, $2$D
and $3$D.
\end{abstract}


\maketitle

The density of states in a mesoscopic sample
is very important to understand mesoscopic transport
phenomena and also thermodynamic properties. 
It is believed that the properties of a mesoscopic
sample, connected to leads, 
can be formalized in terms of the scattering matrix
\cite{butpram02,hac}. The Friedel-sum-rule (FSR)
relates the density of states (DOS) to the scattering matrix and can be
stated as
$\theta_f (E_{2})-\theta_f (E_{1})\approx \pi N(E_{2},E_{1})$ \cite{fri52}.
In $1$D, $2$D and $3$D, the equality being approximate is almost exact in
the WKB regime where generally transport occurs. Here $N(E_{2},E_{1})$ is
the variation in the number of states in the energy interval
$[E_{1},E_{2}]$ due to the scatterer and
$\theta_f (E) = \frac{1}{2} \sum_j \xi_{j} = \frac{1}{2i}
ln (det [S] )$ \cite{lan61}.
$S$ is the scattering matrix and
$e^{i\xi_{j}}$, j=1,2,....n are the n eigenvalues of the unitary matrix
$S$.  In differential form the FSR can also be stated as
 \begin{equation}
 \frac{\partial}{\partial E}\theta_f(E) = \frac{1}{2i}
 \frac{\partial}{\partial E} ln (det [S] ) 
 \approx \pi [\rho(E) - \rho_0(E)] \hspace{.2cm} ,
 \label{eq:difFSR}
 \end{equation}
 where $[\rho(E) - \rho_0(E)]$ is the variation of the DOS or the
difference in the DOS due to the presence of the scatterer. $\rho(E)$ and
$\rho_0(E)$ can be found by integrating the local DOS (LDOS)
$\rho(x,y,z,E)$ and $\rho_0(x,y,z,E)$.

In 1994, Ref.\cite{butpram02} explored a relation between LDOS and the
scattering matrix.  We refer to it as the B{\"u}ttiker-Thomas-Pretre (BTP) 
sum rule which could be exact but is of limited practical value as far 
as the global DOS is concerned, because of
the following reasons.  First of all
if one wants to integrate the LDOS to find the
DOS, then it is very cumbersome and may not be at all possible for
complicated potentials.  Secondly, in this local formalism, one has to
take the derivatives of the S matrix elements with respect to the local
potential and in some cases one encounters problems in including the
non-local effects and non-local disturbances in the LDOS.  Thirdly, if one
wants to find the change in the S matrix due to an infinitesimal change in
the local potential, then one has to know the change in the local
wave-function due to that infinitesimal change in the local potential and
then use the transfer matrix multiplication method or any other equivalent
method to find the change in the S matrix.  So then integrating the LDOS
to find the DOS is just equivalent to integrating the wavefunction to find
the DOS. Another such relation between the LDOS and the scattering
matrix is given by Titov and Schomerus \cite{tit}.
Whereas, the power of the FSR lies in the fact that it
gives this integrated quantity straightaway,
without having to tamper the internal structure of the potential, but just
by infinitesimally changing the energy of an external probing particle.
Although, the BTP sum rule is of use in certain cases, it is
definitely not a competitor of or a substitute for the FSR. No
doubt, even the authors of Ref. [1], much after the proposition of the BTP
sum rule in 1994, refers to Eq. 1 as the FSR, and are worried about to
what extent it is valid \cite{tan99,yeybutt}.

The scattering matrix of an impurity in a
quantum wire have very unusual features that were not realized until very
recently \cite{bag90,deo98}.
In a single channel quantum wire, in the presence of a
single attractive impurity, taken as a negative delta-function potential,
the transmission probability can go to zero \cite{bag90} for some finite
energy of the incident electron. At the corresponding energy, the
scattering phase-shift shows a discontinuous jump(slip) by $\pi$
\cite{deo98}. It was shown that in the single channel case the Friedel
phase $\theta_f$ is not affected by the discontinuous phase drops
\cite{lee99,tan99}. 

In the multichannel case, when the unitarity of a particular channel is
not present and the electron can escape to a different channel, the
transmission zeroes are replaced by minima and the discontinuous phase
slip by $\pi$ are replaced by continuous and less than $\pi$ phase
drops\cite{my1}. It was also shown that in the multi channel
case too the Friedel phase $\theta_f$ is not affected by these continuous
phase drops\cite{my1}. However, $\frac{\partial \theta_f(E)}{\partial E}$
may not bear any resemblance to $\pi[\rho(E)-\rho_0(E)]$ \cite{my1}. 

In this work we intend to study the FSR
(Eq.~(\ref{eq:difFSR})) for a single channel quantum wire with a 
delta-function impurity.
The single channel case being the most
important because it is in this regime that one can really control the
quantum interference effects and use them to build mesoscopic
devices\cite{datta} and point impurities like the delta
function potential are always present. 
Besides, to study the transport across a quantum
dot connected to two ideal leads on two sides, most theoretical works
model the dot by a single bound state at the site of the dot as
the Coulomb blockade makes the other levels of the dot to be very
far away. As an attractive delta potential is capable of creating such a
single bound state, it was used in Ref. \cite{deo98} to explain the Fano
resonances in quantum dots and the unusual features of the scattering
phase shift observed across the quantum dot. Besides, the present study
provides general understandings, and we also understand the system and the
results of Ref. \cite{tan99} better. 

Scattering due
to a spherical defect leads to asymptotic wave-functions that are plane
waves and whose radial part is of the form
$\psi_{k,l}(r) \hspace{.1cm}\sim \hspace{.1cm}\frac{1}{r} \hspace{.2cm}
Sin (kr-\frac{l\pi}{2}+\eta_l) \hspace{.1cm}$ \cite{ziman}.
Here, $k$ is the wave-number, $l$ is the angular momentum quantum number
and $\eta_l$ is the phase shift due to scattering. According to Friedel
\cite{ziman},
to approximately count the number of states created by the impurity,
consider a large sphere of radius $R$ with
a defect at the center, and 
impose a condition
$\psi_{k,l}(R) = 0.$
Thus we obtain
$kR - \frac{l\pi}{2} + \eta_l = n\pi  \hspace{.2cm}$. 
The states $k$ thus obtained are real and not complex as
it should be.
Generally, scattering states have open boundary
conditions that lead to their characterization in terms of complex
energies. The imaginary part of their complex energy is called self energy
and it arises because of the fact that the states can leak out to infinity
and get absorbed by some detector there irrespective of boundary
conditions \cite{datta}. In this
case also the interaction of the states within the large sphere of radius
$R$ with the region outside the sphere will lead to a self energy. 
In absence of the scatterer, the scattering phase shift $\eta_l = 0$ and
thus one gets Eq.~(\ref{eq:difFSR}) \cite{ziman}.

A more rigorous derivation, including the self energy is given
by Buttiker et al 
\cite{yeybutt}. Earlier treatment of Dashen, Ma and Bernstein
\cite{das} and that of Avishai and Band \cite{avi} are 
also reviewed by these authors, in some of their
works like Refs. \cite{butpram02,tan99,gas}. These treatments
are only valid for large systems. Souma and Suzuki \cite{sou}
provides a straight
forward extension of the work by Avishai and Band to quantum wires and
also suffers from the same drawback.
Ref. \cite{yeybutt} gives 
 $$\frac{\partial }{\partial E}\theta_f(E) +{\it Im}\hspace{.1cm}
 Tr \hspace{.1cm} \hat G_a\frac{\partial \hat\Sigma ^a}{\partial E} =
 {\it Im}\hspace{.1cm}
 Tr \hspace{.1cm}{\hat G}^a $$
 \begin{equation}
 = \pi \left[\rho(E) - \rho_0(E)\right] \hspace{.2cm}.
 \label{eq:buttfsr}
 \end{equation}
 $\mbox{Here} \hspace{.5cm} {\hat G}^a = \left[ E - {\hat H}_{system} - {\hat
 \Sigma }^a(E)\right]^{-1} $ ,
 is the advanced Green's function and ${\hat \Sigma}^a$ is the
corresponding self energy.
Note that $\hat G^a$ is the advanced Greens function for the system
alone, where the modifications in the system due to the presence of the
leads is included. Hence apart from this $Im$ $Tr \hat G^a$ which is equal
to the integrated disturbance in the LDOS created by the impurity, i.e.,
$\pi [\rho(E) - \rho_0(E)]$, there will be some disturbance in the LDOS in
the leads which will depend on the Greens function of the lead and how it
is affected by the system. This contribution is not important and also it
gets screened away very easily as it is very small and the leads being
ideal, carrier concentration is very high in the leads (this is often
referred to as non-polarisable leads). 
The only assumption required to get Eq.(\ref{eq:difFSR}) is to neglect
$\partial {\hat \Sigma}^a \over \partial E$ i.e. the energy dependence of
the self energy.
Now \cite{datta}
 \begin{equation}
 \frac{\partial}{\partial E} {\hat \Sigma}^a = \frac{\partial}{\partial E}
 \left[\tau_p g_p^a
 \tau_p^\dagger\right] \hspace{.2cm} ,
 \label{eq:selftrans}
 \end{equation}
$\tau_p$ is the matrix element which couples the
leads to the sample.
In 1D, 2D and 3D 
$\tau_p$ 
is independent of energy
in the WKB regime.
$g_a^p$ being the local Greens function of a semi-infinite
ideal wire, its energy dependence is not very important
and thus in 1D, 2D
and 3D, in the WKB regime, FSR~[(\ref{eq:difFSR})] is
almost exact. There will be violations in the non-WKB regime, but since
transport occurs in WKB regime, these violations are not important. 

 The scattering matrix $S$ for this single channel quasi-one-dimensional
(Q1D) system is
 \begin{equation}
 S = \left(\begin{array}{cc}
 \displaystyle \tilde r_{11} &
 \displaystyle \tilde t_{11} \\
 \displaystyle \tilde t_{11} &
 \displaystyle \tilde r_{11}
 \end{array}\right) \hspace{.2cm}.
 \label{eq:scatmat}
 \end{equation}

Bagwell \cite{bag90}
has obtained
 \begin{equation}
 \tilde t_{11} = 1+\frac{-i\frac{\Gamma_{11}}{2k_1}}
 {1+\sum_{n>1}\frac{\Gamma_{nn}}{2\kappa_n}+i\frac{\Gamma_{11}}{2k_1}},
 \label{eq:t11}
 \end{equation}
 \begin{equation}
 and \,\,\tilde r_{11} = - 1 + \tilde t_{11} ,
 \end{equation}
 $\mbox{where,}\hspace{.5cm}\Gamma_{nm}=\frac{2m_e \gamma}{\hbar^2}
Sin[\frac{n\pi}{w}(y_i+\frac{w}{2})]
Sin[\frac{m\pi}{w}(y_i+\frac{w}{2})],$
$\kappa_n=\sqrt{\frac{2m_e}{\hbar^2}(E_n-E)}$,
$k_n=\sqrt{\frac{2m_e}{\hbar^2}(E-E_n)}$, $E_n={\hbar^2 \over 2m} {n^2 \pi^2
\over w^2}$, $\gamma$ is the strength of the delta potential placed
at a distance $y_i$ from the center of the quantum wire of width $w$.
We find that a similar expression exists for the transition amplitude
from the propagating mode ($n$=1) to the nth evanescent mode, given by
 \begin{equation}
 t_{1n}=\frac{-\frac{\Gamma_{1n}}{2\kappa_n}}{1+\sum_{n>1}\frac{\Gamma_{nn}}
 {2\kappa_n}+i\frac{\Gamma_{11}}{2k_1}}
 \label{eq:t1n}
 \end{equation}
 When the impurity potential is positive it
can only support scattering states. However when the impurity potential is
negative, it can also support bound states, apart from the scattering
states. For each n we get a sub-band
of scattering states ($E$ as a function of $k_n$). Similarly we get a
bound state for each n, that are solutions to \cite{bag90}
 \begin{equation}
 1+\sum_{m=n}^{\infty} \frac{\Gamma_{mm}}{2\kappa_m}=0\hspace{.5cm}.
 \label{eq:trubnd}
 \end{equation}
 For n=1 we get a true bound state.
The bound state for n=2 may or may not be a true bound state. If the
impurity potential is such that the solution to Eq.(\ref{eq:trubnd})
lie in the energy range where $n$=1 channel is propagating, 
then this bound state for
$n=2$, is degenerate with n=1 scattering state and it becomes a
quasi-bound state.

We have also calculated to find 
 \begin{eqnarray}
 [\rho(E) - \rho_0(E)]_{total}& = & \sum_p \frac{2|{\tilde r_{pp}}|}{hv_p}
 \int_{-\infty}^{\infty} dx\hspace{.2cm} Cos(2k_px+\eta_p)\nonumber\\
 &+&\sum_p \frac{2}{hv_p}\hspace{.2cm}\sum_e\frac{\mid t_{pe}\mid^2}{\kappa_e}
 \hspace{.2cm}.
 \label{eq:sysDOS}
 \end{eqnarray}
 Here $v_p=\hbar k_p/m$. 
$\sum_p$ denotes sum over all propagating modes and $\sum_e$ denotes sum
over all evanescent modes. The 1st term on the R.H.S. is basically 
integrated change in the LDOS in the leads. 
Since the delta function potential is a point impurity, the
integrated LDOS in the leads extends from $-\infty$ to $\infty$. One can
do the integration to find $\int_{-\infty}^{\infty} dx Cos(2k_p x+\eta_p)
= \pi Cos(\eta_p) \delta(k_p)$.  So 
it is zero unless the
quasi-bound state coincides with $k_p=0$.
In the case of extended impurities
one can see that this term gives an
unimportant small contribution that does not change with energy.
Also the carrier concentration in the
leads is normally large enough to screen away a small oscillatory LDOS
completely. So the relevant quantity that appears in FSR is
 \begin{equation}
 \rho(E) - \rho_0(E) = \sum_p\frac{2}{hv_p}\hspace{.2cm}\sum_e \frac{
 \mid t_{pe}\mid^2}{\kappa_e}  ,
 \label{eq:rhoexpr}
 \end{equation}
 This is actually
the integrated local DOS around the impurity site and decaying away from
the impurity site all the way up to $\pm \infty$.  In
the regime of single propagating channel $p$ can take only one value, i.e.
$p=1$ and $e=2, 3, ...\infty$.
Thus we can independently calculate both sides of
Eq.~(\ref{eq:difFSR}) starting from 1st principles, where we do not
have to throw away dispersive behavior or energy dependence of self
energy.

We first present below a discussion and definition of the WKB regime
for a Q1D system, because it is an interesting subject on its own.
When the
incident electron propagates in a potential where the wavefunction 
changes very slowly in space then
very little reflected wave is generated and that is taken to be the WKB
regime~\cite{merzbook}. So a delta function potential in one-dimension
(1D) has a WKB
regime at higher energies, 
when the reflection probability is very small.  In the inset of
Fig.1, where we plot $|\tilde r_{11}|^2$ versus incident energy
we find that there are three regimes. One is to the left of point $P_1$
where $|\tilde r_{11}|^2$ is large and also strongly energy dependent. 
The other is between the points $P_1$ and $Q_1$ where ${\hbar^2 k_1^2
\over 2m_e}>>\gamma$. These two regimes can be seen in 1D scattering 
(e.g., a delta function potential in 1D) and
are the non-WKB and WKB regimes, respectively.  The third regime is to the
right of the point $Q_1$, where again $|\tilde r_{11}|^2$ is very small
and is hence a WKB regime, but the energy dependence of $|\tilde
r_{11}|^2$ is very large. Such a regime cannot be seen in 1D and is a
specialty of Q1D.  So the energies that lie to the left of $P_1$ is 
the non-WKB regime, where
the electron feels the potential very strongly and is almost entirely
reflected back. Energies to the right of the
point $P_1$ correspond to the WKB limit.  Although, the system considered
here is a Q1D system, corresponding to a scatterer in Q1D, there is an
energy dependent scatterer in 1D \cite{hua, tek, legdeo}. 
The bound states and scattering states of these two
potentials are identical and this is an exact correspondence, valid in all
regimes, quantum or semi-classical.  
And so when the reflection probability is
small in Q1D, it is also small in the corresponding 1D potential. Then all
the notions and results of WKB regime that we are familiar with in
1D are also true in Q1D.

In Fig.1 we find a large deviation of $\pi [\rho(E)-\rho_0(E)]$
(dotted curve) from $d \theta_f \over dE$ (solid curve) at energies in the
non-WKB regime (left of $P_1$). This is similar to what is seen in 1D, 2D
or 3D. In the WKB regime, that is to the right of the point $P_1$,
although $\mid \tilde r_{11}\mid ^2$ is very small, its energy dependence
is not as negligible as that of a potential in $1$D (eg, a delta function
potential in $1$D or a square well in $1$D). Energy dependence of $|\tilde
r_{11}|^2$ automatically implies energy dependence of $\tau_p$ (or
$\hat \Sigma^a$), i.e.,
dispersive behavior. So there is an appreciable difference between
$\pi[\rho(E)-\rho_0(E)]$ and $\frac{d\theta_f}{dE}$.

Let us now analytically analyse the
curves (solid and dashed) to the right of $Q_1$. In this region
$\kappa_2\rightarrow 0$.  From
Eq.(\ref{eq:rhoexpr}) we find that only the 1st term in the series is
relevant. That is
 \begin{equation}
 \pi \left[ \rho(E)-\rho_0(E)\right]_{\kappa_2 \rightarrow 0}
 \mbox{diverges as}
 \left[ \frac{2\pi}{hv_1} \frac{\Gamma_{11}}{\Gamma_{22}}
 \frac{1}{\kappa_2}
 \right]_{\kappa_2 \rightarrow 0}
 \label{eq:rholtK}
 \end{equation}
 Thus the strong energy dependence in the energy beyond $Q_1$ is
due to the rapid population of the second subband through evanescent
modes, as it approaches its propagating threshold.
Hence, unlike the Fano resonance, this is not a quantum 
interference effect.
This is like the Van Hove singularity at the band edge.
Similarly one can find
 \begin{equation}
 \left[ \frac{d\theta_f}{dE}\right]_{\kappa_2 \rightarrow 0}=
 \left[\frac{d}{dE} arg(\tilde t_{11})\right]_{\kappa_2 \rightarrow 0}
 \label{eq:thft11}
 \end{equation}
 diverges identically.
Note that although $arg(\tilde t_{11})$ can have a discontinuity, the
derivative exists at all energies. Essentially the right derivative and
left derivative is the same at the discontinuity.  Hence we
prove FSR is exact as $\kappa_2 \rightarrow 0$.
This is understood
when we note that when $\kappa_2 \rightarrow 0$,
$|\tilde r_{11}|^2$ goes to zero at the band
edge \cite{bag90}.  Also it is known that when
$\tilde r_{11}=0$ then $\tau_p$ maximizes \cite{datta}, and energy
dependence of $g_p^a$ being negligible, $d \hat\Sigma ^a \over dE$=0 .
Thus all the deviating terms being zero,
the FSR is valid around the diverging DOS at the band edge.

For strong negative potentials, such that the bound state for
$n=2$ is below the propagating threshold of $n=1$, the curves look similar
to that in Fig. 1.
For the negative $\delta$ function potential with the bound state 
for $n=2$, in the
propagating regime of $n=1$, we have plotted the two sides of
Eq.(\ref{eq:difFSR}) in Fig.2.  $|\tilde t_{11}|^2$ is shown in
the inset. Note that $|\tilde t_{11}|^2$ shows that at the point P the
system is in extreme non-WKB regime where $|\tilde t_{11}|^2$ goes to
zero.  
At this energy there is a quasi-bound state and there is strong
energy dependence of scattering matrix elements as well as self energy.
According to earlier stated results \cite{butpram02}, 
there should be violation of
FSR here.
The peak in $\pi[\rho(E)-\rho_0(E)]$ at P occurs due to the quasi-bound
state. We also see that at this very point P there is an exact agreement
between L.H.S. \& R.H.S. of Eq.(\ref{eq:difFSR}).  This can be even
verified analytically.  Substituting the bound state condition given in
Eq.(\ref{eq:trubnd}) into $\frac{d\theta_f}{dE}$ as well as in 
$\pi[\rho(E)-\rho_0(E)]$ separately, we get
 \begin{equation}
 \frac{d\theta_f}{dE} = \frac{m_e k_1}{\hbar^2}\frac{1}{\Gamma_{11}}\sum_{n>1}
 \frac{\Gamma_{nn}}{\kappa_n^3}
 =\pi[\rho(E)-\rho_0(E)]
 \label{eq:frthdeBS}
 \end{equation}
 This agreement between $\frac{d\theta_f}{dE}$ and
$\pi[\rho(E)-\rho_0(E)]$ was argued to be equal for the case of a stub in
ref \cite{tan99}, at the transmission zero where the $\frac{\partial
\hat\Sigma ^a}{\partial E}$ term in Eq.(\ref{eq:buttfsr})
was dropped from the very
beginning\cite{tan99}. Dropping the energy dependence of $\hat \Sigma^a$ 
in non-WKB regime and
verifying the validity of FSR is rather meaningless, as the violations do
come from the energy dependence of $\hat \Sigma^a$.  
Even after including these
terms we get exact agreement at the transmission zero for the negative
$\delta$ function potential in a quantum wire, although it
is in extreme non-WKB regime.  The reasons are as follows. At the
transmission zero, since there is a quasi-bound state, $\hat\Sigma ^a$
becomes minimum and $\frac{\partial \hat\Sigma ^a}{\partial E}=0$.
All these arguments are also true for the stub.

The multichannel case was analysed in Ref.\cite{my1}. When two
modes are propagating, the bound state coming from the 3rd sub-band, can
be degenerate with the propagating channels. At that point there is no
agreement between $\frac{d\theta_f}{dE}$ and $\pi [\rho(E)-\rho_0(E)]$.
While $\pi [\rho(E)-\rho_0(E)]$ is not only positive definite, it also has
a sharp peak meaning enhanced DOS at the energy corresponding to the bound
state. The peak is completely missing from $\frac{d\theta_f}{dE}$ and
$\frac{d\theta_f}{dE}$ turns out to be negative. Thus the disagreement is
not just a quantitative one, but is a
qualitative one.

Thus the purpose of this work was to verify FSR in single channel
quantum wires in the presence or absence of Fano
resonances. Fano resonance is a very general feature
of quantum wires. At the Fano resonance all the quantities are
strongly wave vector dependent as it is a purely quantum interference
effect. Never the less, FSR is exact at the
Fano resonance. This is contrary to the known fact that FSR is valid
in semiclassical regimes where there is no strong dependence on wave
vector. The exact agreement of the FSR in spite of the strong wave vector
dependence is due to the fact that at the Fano resonance there is a
quasi bound state that leads to a minimum in the self energy.
Away from this quasi bound state there are
strong violations. These are true for any negative potential in Q1D and
the potential considered here and the associated calculations
make this clear. For positive as well as negative delta function
potentials, there is also strong wave vector dependence, close
to the upper band edge of single channel propagation. 
This is due to the rapid population of the first evanescent
mode at its propagation threshold
and does not depend on the
existence of Fano resonance.
$d\Sigma/dE$=0 here because of the perfect transmission at the
band edge, and hence the agreement in FSR.
So the former case of agreement in the peak is an agreement in
purely quantum regime, while that in the case of the latter peak
is in the semiclassical regime. Away from the peaks there is
always violation.
It may be interesting to work out
some extended potentials in
Q1D \cite{stub}.

One of the authors(S.B.) gratefully thanks Prof. Binayak Dutta Roy and
Debasish Chaudhuri for useful discussions.

\centerline{\bf Figure Captions}

\noindent Fig. 1. The
dashed curve gives $\pi(\rho-\rho_0)$ and the solid curve gives
$\frac{d}{dE}(-.5i\hspace{.1cm}ln\hspace{.1cm} Det[S])$. Both the
functions are plotted versus $EW^2$ using $y_i=.21W$ and
$\gamma=1$. In the inset the corresponding $|\tilde
r_{11}|^2$ is plotted. 
We have considered $500$ evanescent modes.

\noindent Fig. 2.  The solid
curve gives $\pi(\rho-\rho_0)$ and the dashed curve gives
$\frac{d}{dE}(-.5i ln Det[S])$. Both the functions are plotted versus
$EW^2$ using $y_i=.21W$ and $\gamma=-1.5$.
For this value of $\gamma$ there is a quasi-bound state at $EW^2=36.1022$.
In the inset the corresponding $|\tilde t_{11}|^2$ is plotted.
We have considered $500$ evanescent modes.


\begin{thebibliography}{99}
\bibitem{stub} We have found that if we take two delta function potentials
in a single channel quantum wire and place two of them close to each other,
then for very large strengths of the two potentials, the system becomes
almost reflectionless. This is very counter intuitive as large potentials
are expected to give large reflections. It happens because strong potential
couples the evanescent modes very strongly to the propagating mode at the
positions of the delta potentials, and an incident electron is thrown
into the evanescent channels rather than reflected.
And since the two potentials are close
to each other, the electron can easily tunnel from one potential site to the
other and finally go to the transmission channel, again due to the large
coupling between the evanescent and transmission channels.
As far as FSR is concerned, 
qualitative results will be the same. At the quasi-bound states,
the self energy has a minima, which means $d\Sigma/dE$=0, and so the FSR
will be exact at these points. There will be violations away from these
points, the extent of which has to be seen through explicit calculations.
\end{thebibliography}
\end{document}